\documentclass[iop]{emulateapj}
\usepackage[utf8]{inputenc}
\usepackage{gensymb}
\usepackage{units}
\usepackage{hyperref}
\usepackage{graphicx}
\usepackage{multirow}
\usepackage{color}
\usepackage{lineno}

\begin{document}

\newcommand{\casa}{\texttt{CASA }}
\newcommand{\emcee}{\texttt{emcee }}
\newcommand{\galario}{\texttt{galario }}

\bibliographystyle{apj} 

\title{Multiwavelength Vertical Structure in the AU Mic Debris Disk: Characterizing the Collisional Cascade}

\author{David Vizgan\altaffilmark{1,2,3,4}}

\author{A. Meredith Hughes\altaffilmark{1}}

\author{Evan S. Carter\altaffilmark{1}}

\author{Kevin M. Flaherty\altaffilmark{5}}

\author{Margaret Pan\altaffilmark{6}}

\author{Eugene Chiang\altaffilmark{7,8}}

\author{Hilke Schlichting\altaffilmark{9}}

\author{David J. Wilner\altaffilmark{6}}

\author{Sean M. Andrews\altaffilmark{6}}

\author{John M. Carpenter\altaffilmark{10}}

\author{Attila Mo\'{o}r\altaffilmark{11}}

\author{Meredith A. MacGregor\altaffilmark{12}}

\altaffiltext{1}{Astronomy Department and Van Vleck Observatory, Wesleyan University, 96 Foss Hill Drive, Middletown, CT 06459, USA}
\altaffiltext{2}{Department of Astronomy, University of Illinois at Urbana-Champaign, 1002 West Green St., Urbana, IL 61801, USA}
\altaffiltext{3}{The Cosmic Dawn Center}
\altaffiltext{4}{National Space Institute, DTU Space,  Technical University of Denmark, 2800 Kgs.~Lyngby, Denmark}
\altaffiltext{5}{Department of Astronomy and Department of Physics, Williams College, Williamstown, MA 01267, USA}
\altaffiltext{6}{Center for Astrophysics \textbar\ Harvard \& Smithsonian, 60 Garden Street, Cambridge, MA 02138, USA}
\altaffiltext{7}{Department of Astronomy, University of California at Berkeley, Campbell Hall, Berkeley, CA 94720-3411, USA}
\altaffiltext{8}{Department of Earth and Planetary Science, University of California at Berkeley, McCone Hall, Berkeley, CA 94720-3411, USA}
\altaffiltext{9}{University of California, Los Angeles, Los Angeles, CA, USA}
\altaffiltext{10}{Joint ALMA Observatory, Alonso de C\'{o}rdova 3107 Vitacura, Santiago, Chile}
\altaffiltext{11}{Konkoly Observatory, Research Centre for Astronomy and Earth Sciences, E\"{o}tv\"{o}s Lor\'{a}nd Research Network (ELKH), Konkoly Thege Mikl\'{o}s \'{u}t 15-17., H-1121 Budapest, Hungary}
\altaffiltext{12}{Department of Astrophysical and Planetary Sciences, University of Colorado, 2000 Colorado Avenue, Boulder, CO 80309, USA}

\begin{abstract}
   

Debris disks are scaled-up analogs of the Kuiper Belt in which dust is generated by collisions between planetesimals. In the ``collisional cascade" model of debris disks, dust lost to radiation pressure and winds is constantly replenished by grinding collisions between planetesimals. The model assumes that collisions are destructive and involve large velocities; this assumption has not been tested beyond our Solar System. We present 0\farcs25 ($\approx$2.4\,au) resolution observations of the $\lambda$ = 450\,$\mu$m dust continuum emission from the debris disk around the nearby M dwarf AU Microscopii with the Atacama Large Millimeter/submillimeter Array. We use parametric models to describe the disk structure, and an MCMC algorithm to explore the posterior distributions of the model parameters; we fit the structure of the disk to both our data and archival $\lambda = 1.3$\,mm data \citep{daley2019}, from which we obtain two aspect ratio measurements at 1.3 mm ($h_{1300}$ = 0.025$^{+0.008}_{-0.002}$) and at 450 $\mu$m ($h_{450}$ = 0.019$^{+0.006}_{-0.001}$), as well as the grain size distribution index $q =$ 3.03 $\pm$ 0.02. Contextualizing our aspect ratio measurements within the modeling framework laid out in \citet{pan2012}, we derive a power law index of velocity dispersion as a function of grain size $p =  0.28 \pm 0.06$ for the AU Mic debris disk. This result implies that smaller bodies are more easily disrupted than larger bodies by collisions, which is inconsistent with the strength regime usually assumed for such small bodies. Possible explanations for this discrepancy are discussed.

\end{abstract}


\section{Introduction}\label{chap1}

Debris disks form from dust produced by collisions between planetesimals during the late stages of the planet formation process \citep[][and references therein]{hughes2018}. They are detected around 15-20\% of main sequence stars, although since current sensitivity limits are at least an order of magnitude above what would be required to detect our Solar System's Kuiper Belt, this incidence should be considered a lower limit \citep{matthews2014,montesinos2016,sibthorpe2018}. Debris disks are often described as ``left over'' from planet formation, but this description downplays the dynamic, evolving nature of the bodies within the disk. While stellar processes, such as radiation pressure and stellar winds, remove dust and other small particles from a debris disk \citep[][]{wyatt2002, chen2006}, collisional processes continually replenish the dust \citep{wyatt2008, matthews2014, hughes2018}. The combination of processes is known as a ``collisional cascade'' \citep{dohnanyi1969}, where the objects in the disks, from the larger asteroids and comets to the smaller dust grains, undergo inelastic collisions. These interactions fragment larger planetesimals, transfer mass to smaller planetesimals, and replenish the dust which is then removed by radiation pressure and stellar winds. 
 
The theoretical description of the collisional cascade is parameterized by two power laws: the velocity dispersion $v$ as a function of grain size $a$, which has power law index $p$, and the grain size distribution $\frac{dN}{da}$ as a function of grain size, which has power law index $q$:
 \begin{equation}
    v(a) \propto a^{p}
    \label{eq:p}
 \end{equation}
\begin{equation}
    \frac{d N}{d a} \propto a^{-q}
    \label{eq:q}
\end{equation}
The grain size distribution index $q$ has been shown to be remarkably insensitive to parameters which describe the fragmentation process e.g. size and number of the largest fragments, size distribution of fragments \citep[][]{williams1994} and found to be constant in the case of self-similar fragmentation outcomes \citep{tanaka1996}.

The parameters $p$ and $q$ are described within the context of a ``gravity'' and ``strength'' regime, referring to whether gravitational forces or internal strengths of grains determine the outcome of collisions between bodies \citep[][]{benz1999}; typically, small objects are bound by material strength, while larger planetesimals are bound by self-gravity \citep[][]{lohne2008}. 

 This parametric framework has been applied to the Kuiper Belt to describe its collisional evolution \citep[][]{bernstein2004, pan2005} and more generally to the collisional evolution of debris disks \citep[][]{lohne2008, wyatt2008}. Building upon this work, \cite{pan2012} investigated the grain-size-dependent velocity dispersion within the collisional cascade framework and found self-consistent steady state solutions for $p$ and $q$ within both the strength and gravity regimes. They pointed out that observing the scale heights of debris disks could constrain the mass and number of large bodies stirring the cascade as well as the internal strengths of colliding bodies. 
 
 The vertical structure of a debris disk encodes information about the dynamical state of the collisional cascade.  The scale height $H(r)$ is typically the standard deviation of the Gaussian profile of dust density as a function of height above the midplane. It is defined at a given radius, and is generally assumed to increase linearly with distance $r$ from the star, with a constant ``aspect ratio" $h$, such that $h = H(r) / r$.  However, the observed scale height varies as a function of wavelength, and factors other than the dynamical state of the collisional cascade can influence the measured scale height, particularly at short (optical and infrared, or OIR) wavelengths. 
 
\cite{thebault2009} showed that all debris disks have a natural minimum scale height at OIR wavelengths due to radiation pressure, stellar winds, and mutual grain collisions \citep[which can also be altered by the gas content; see ][]{olofsson2022}. Measurements of the scale height at OIR wavelengths therefore cannot be used to probe the dynamical conditions within a disk. However, millimeter-wavelength observations trace larger grains that are resistant to the effects of radiation pressure and stellar winds \citep[for review see][and references therein]{hughes2018}, which means that measurements of the scale height at millimeter wavelengths can be used to measure the dynamical excitation of the dust grains. 

Converting a scale height measurement into a velocity dispersion involves assuming an equipartition between eccentricity and inclination induced by dynamical interactions between dust grains and planetesimals in the disk, so that orbital inclination $i$ is $\sim$ $\sqrt{2}h$ for a sufficiently small $h$ \citep[][]{wyatt2002}. This aspect ratio, when measured at a particular wavelength, is directly proportional to the velocity dispersion for grains with sizes comparable to the wavelength of observation (see \S 4.2), which is, in turn, proportional to $a^p$. Thus, scale height measurements of different grain size populations (i.e. at different wavelengths) within a debris disk can describe the velocity dispersion of those bodies across a range of grain sizes, allowing us to measure the velocity dispersion power law index $p$.

Within the millimeter-wavelength regime, the scale height can still vary with wavelength, since different wavelengths of light can probe different dust grain sizes.  This feature results from a balance between the grain size distribution, which tends to tilt heavily towards smaller grains that dominate the surface area, and the falloff of grain emission efficiency with wavelength for wavelengths longer than the size of the dust grain.  Taken together, these two trends result in the rule of thumb that the smallest grain that can emit efficiently will dominate the observation, and that the smallest grain that can emit efficiently is one that is comparable in size to the wavelength of light.  With observations of a disk at multiple wavelengths, therefore, we can derive values for $p$ and $q$ and ultimately parameterize the collisional cascade of a disk via interpretation within the framework described in \cite{pan2012}.
 
The power-law index for grain size distribution, $q$, has been inferred via modeling for many debris disks beyond our own Solar System \citep[e.g.][]{lohne2012, ricci2015, macgregor2016, marshall2017, white2018, hengst2020, arnold2021, norfolk2021}; nearly all of these systems are consistent with the theoretical steady-state grain size distribution index $q = 3.5$ derived by \cite{dohnanyi1969}. Generally, a velocity dispersion of 1 km/s is prescribed to all bodies within the Kuiper Belt \citep[][]{leinhardt2008}, which suggests $p=0$ locally \citep[as assumed by][]{dohnanyi1969}; it is important to note that non-negligible differences in collisional velocities have been measured in various dynamical populations of Kuiper Belt objects \citep[e.g.][]{abedin2021}. Nonetheless, measurements of $p$ have never been made outside of the Solar System. 

AU Microscopii, also known in the literature as GJ 803 and HD 197481, is a prime target for the first measurement of $p$ via multiple resolved scale height measurements at widely separated millimeter wavelengths.  It is an M3IVe star \citep[][]{daley2019} located $9.725 \pm 0.005$\,pc away from our Solar System \citep{gaia2018} that hosts one of the biggest, brightest, and most edge-on debris disks in the night sky. It is a member of the $\beta$ Pictoris Moving Group and has an age of 22 $\pm$ 3 million years \citep[][]{mamajek2014}. AU Mic has a mass of 0.50 $\pm$ 0.03 M$_{\rm \odot}$ \citep[][]{plavchan2020}, a luminosity  of 0.09 L$_{\rm \odot}$, and an effective temperature of 3700 $\pm$ 100 K \citep[][]{plavchan2009}. AU Mic is also home to two Hot Neptunes at separations of 0.0645 $\pm$ 0.0013 and 0.1101 $\pm$ 0.0022\,au from the star \citep{plavchan2020, martioli2020}, both of which may be able to induce radio emission in the star's corona \citep{kavanagh2021}. There is also some preliminary evidence for a possible third planet orbiting between AU Mic b and c \citep[][]{wittrock2022}. 

AU Mic has been observed across the electromagnetic spectrum, from scattered light \citep{liu2004b,krist2005,augereau2006,graham2007,schneider2014,schuppler2015,grady2020}, to the infrared \citep[e.g.][]{moshir1990, liu2004a, matthews2015}, to millimeter wavelengths \citep{wilner2012,macgregor2013,macgregor2016,daley2019}. The star's violent flaring activity has even been detected in the ultraviolet \citep{robinson2001, redfield2002, roberge2005} and in x-ray \citep{pye2015} as well as in the millimeter \citep{daley2019,macgregor2020}. \cite{boccaletti2015} detected fast-moving features in scattered light within the debris disk. These features, sustained over several years \citep[][]{boccaletti2018}, were attributed by \cite{chiang2017} to dust clouds produced by ``dust avalanches'' blown out by the star's stellar wind, and attributed by \cite{sezestre2017} to sequential dust releases from an unseen parent body orbiting AU Mic. Recent modelling work of AU Mic's stellar wind predicts violent space weather and extreme coronal mass ejections from the star \citep[][]{alvarado2022}; the star's recently measured magnetic field \citep{kochukhov2020, klein2021} might also be responsible for the turbulent and violent coronal ejections \citep[e.g.][]{kavanagh2021} from the star. 

The AU Microscopii debris disk is almost perfectly edge-on, with an inclination very close to 90$\degree$ \citep[][and references therein]{daley2019}. Because the AU Mic debris disk is already resolved vertically at one wavelength, resolving the vertical structure at another wavelength (and therefore another grain size) allows us to parameterize the disk's collisional cascade via calculation of the power law indices $p$ and $q$. \cite{macgregor2016} measured an upper limit on the grain-size distribution power law index $q < 3.31$ for the disk using 9 mm observations from the Very Large Array. \cite{daley2019} used ALMA observations at 1.3 mm to resolve the vertical structure of the disk. Their work measured a disk aspect ratio of $h = 0.031^{+0.005}_{-0.004}$ at 1.3\,mm and used it to infer an upper limit of 1.8\,M$_\earth$ on the total mass of perturbing bodies between radii of $\sim$20-40\,au. 

This work presents new 450\,$\mu$m ALMA observations of the AU Mic debris disk, at 0\farcs25 ($\approx$2.4\,au) resolution -- the same as the 1.3\,mm data from \citet{daley2019}. We model the debris disk structure, combining measurements of the scale height and the flux of the disk at both 1.3\,mm and 450\,$\mu$m to calculate $p$ and $q$ for the AU Mic disk. The observations of the AU Mic debris disk are presented in Section~\ref{chap2}, along with the data reduction and post-processing methods. The main results of the debris disk observations are presented in Section~\ref{chap3}. Our analysis of the visibilities and measurement of the 1.3\,mm / 450\,$\mu$m scale height ratio is presented in Section~\ref{chap4}, and the results of this work are discussed in Section~\ref{chap5}. This work concludes with a brief summary of the main results in Section~\ref{chap6}. 

\section{Observations}\label{chap2}

\subsection{450\,$\mu$m disk observations}
AU Mic was observed with the Atacama Large Millimeter/submillimeter Array (ALMA) in the Atacama Desert, Chile, as part of project 2016.1.00878.S (PI: A.~M.~Hughes). The system was observed in five executions of one scheduling block; once on 2017 April 22, twice on 2017 April 23, and twice on 2017 April 26. The weather during observations was excellent, with precipitable water vapor $<$ 1 mm for all execution blocks. This information, along with beam sizes and calibration sources, is presented in Table \ref{tab:obs}. The observations utilized ALMA's Band 9 ($\lambda = 450$ $\mu$m) receivers and 45 12m antennas. The correlator was configured to maximum continuum sensitivity by employing four spectral windows each with a bandwidth of 1.875 GHz. Each window was divided into 128 channels, each with a width of 15.6 MHz (or $\sim$ 21 km/s). 

The primary beam of ALMA's 12 m antennas is modeled as a 2D Gaussian with a FWHM (full-width half-maximum) equal to 
$1.13 \lambda / D$ (as described in the ALMA Technical Handbook for Cycle 7) with a peak response normalized to unity. Because the width of the primary beam at $\lambda = 450 \ \mu$m is comparable to the width of the disk, the observations were conducted in a three-pointing mosaic, with position offsets of (+3\farcs0, -2\farcs3) and (-3\farcs0, +2\farcs3) in right ascension and declination, centered on the position of AU Mic and oriented along the previously measured position angle of 128.4$\degree$ \citep{macgregor2013}. 


The four spectral windows were tuned to central frequencies of 676, 678, 680, and 682 GHz. The baseline lengths span between 12 m and 455 m; the 455 m baseline traces an angular scale of 0.2" and a spatial scale of 2 au.  Calibration and imaging of the visibilities were conducted via the \texttt{CASA} software package \citep{mcmullin2007}. Standard ALMA reduction scripts were applied to the data; gain calibration was carried out via a combination of water vapor radiometry and on-sky gain calibrators, listed in Table \ref{tab:obs}. Flux calibration was carried out using quasars with fluxes bootstrapped from the ALMA calibrator list. The ALMA technical handbook for Cycle 7 quotes a 20\% systematic flux calibration uncertatinty at a wavelength of 450 $\mu$m. The statistical weights of the visibilities were recalculated and applied to the data by using the variance around each baseline \cite[see][]{flaherty2017}. For computational efficiency, we average the data into spectral bandwidths of 1.875 GHz and within a time interval of 30 seconds.

\begin{table*}[t!]
    \centering
    \resizebox{\textwidth}{!}{
    \begin{tabular}{|c|c|c|c|c|c|}
         \hline
         Parameters & 2017 Apr 22 & 2017 Apr 23 & 2017 Apr 23 & 2017 Apr 26 & 2017 Apr 26 \\
         \hline 
         Precipitable water vapor (mm) & 0.36 & 0.58 & 0.65 & 0.32 & 0.35 \\
         On-source time (min) & 100 & 90 & 90 & 94 & 75 \\
         Flux calibrator & J1924-2914 & J1924-2914 & J1733-1304 & J1924-2914 & J1924-2914 \\
         Bandpass calibrator & J1924-2914 & J1924-2914 & J2232-1143 & J1924-2914 & J2232-1143 \\
         Gain calibrator & J1957-3845 & J1957-3845 & J1957-3845 & J1957-3845 & J1957-3845 \\
         \hline
    \end{tabular}}
    \caption{Observational parameters for the five execution blocks. Forty-five 12 m antennas, spanning baselines of 12 to 455 m, were used for the observations. }
    \label{tab:obs}
\end{table*}

The \texttt{CASA} task \texttt{tclean} was used to image the data using standard Fourier transform methods. A bright submillimeter source --- most likely a galaxy --- was detected to the East of the AU Mic debris disk (we note that the offset of the galaxy from the star is so large that it is not visible within the boundaries of Figure \ref{fig:AUMic}). In order to subtract this galaxy from the visibilities we used a mix of \texttt{CASA} and the Python package \texttt{galario} \citep{tazzari2018}. We first used the \casa task \texttt{tclean} to create a naturally weighted image. We used the elliptical Gaussian fitting tool in the \texttt{CASA} task \texttt{viewer} to fit a 2-D Gaussian at the position of the galaxy. The fit yielded a major axis FWHM = 448 mas, a minor axis FWHM = 365 mas, an integrated flux of 1.59 mJy, a position angle of 175.7$\degree$, as well as a position of $\alpha$ = 20h 45m 10.28s, $\delta$ = -31$\degree$ 20' 33.5". We then used \texttt{galario} to subtract this 2-D Gaussian from each of the twelve visibility files.

 Because the data is a three pointing mosaic, we multiplied each part of our mosaicked model galaxy by a 2D Gaussian with a FWHM $=$ 9\farcs29 at its respective pointing. This primary beam correction was necessary in order to replicate the effects of the primary beam in the ALMA data, which has a different effect on the galaxy for each of three pointings.

\subsection{Stellar variability}

As discussed in the previous section, AU Mic is a young M dwarf, and consequently a highly active star, with a history of flaring at mm wavelengths \citep{macgregor2013, daley2019}. As such, it is important to know whether for these new 450\,$\mu$m data, ALMA observed the star as it flared. \citet{carter2018}, who previously used this same $\lambda = 450 \ \mu$m data to model the disk around AU Mic, used the \casa task \texttt{gaussfit} to fit a Gaussian to the central peak in the image domain for the first and second half of each execution block, finding that the stellar flux varied from a minimum of $1.0 \pm 0.3$\,mJy in the first block to a maximum of $3.0 \pm 0.4$\,mJy in the third block, but did not deviate by more than 3$\sigma$ from the weighted mean of the 10 values over the course of the 5-execution observation. The flaring activity of AU Mic can occur over short (i.e. between 1 second and 1 minute) timescales \citep{macgregor2020}; still, these short stellar flares should not affect the disk modelling substantially due to the high angular resolution of the data.

\section{Results}\label{chap3}

\begin{figure}[h!]
    \centering
    \hspace*{-0.75cm}
    \includegraphics[width=10.5cm]{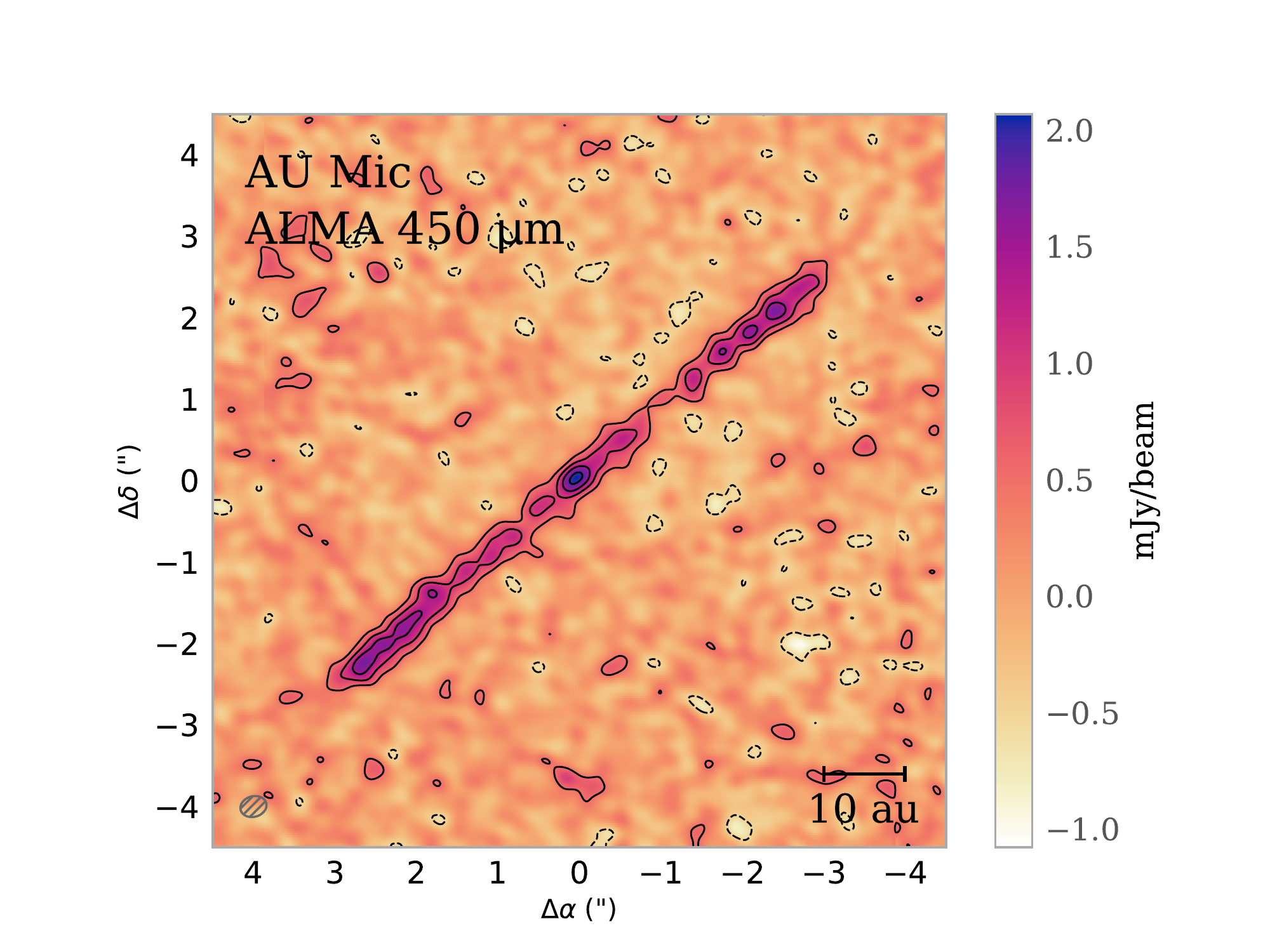}
    \caption{A naturally weighted image of the 450 $\mu$m ALMA observations of the AU Microscopii debris disk. Contours are \mbox{[-2, 2, 4, 6, 8]}$\times \sigma$, where $\sigma = $ the RMS noise $=$ 240 \rm{$\mu$}Jy\,beam\,$^{-1}$. The synthesized beam dimensions, represented by the gray hatched ellipse in the lower left corner, are 0.33" x 0.25" with a position angle of 73.7\degree.}
    \label{fig:AUMic}
\end{figure}

The 450 $\mu$m emission from the disk around AU Mic is shown in Figure \ref{fig:AUMic}. The RMS noise in the image is 240 $\mu$Jy\,beam\,$^{-1}$. By fitting a polygon to the 2$\sigma$ contours of the flux via the \texttt{CASA} task \texttt{viewer}, we measure an integrated flux density of $15.3 \pm 1.1$\,mJy. Fluxes at the position of the star and within the ansae of the disk are also measured using the fitting tool in the \texttt{CASA} task \texttt{viewer}. The peak intensity at the position of the star is $1.6 \pm 0.2$\,mJy\,beam$^{-1}$, the northeast ansa of the disk has a peak intensity of $1.6 \pm 0.2$\,mJy\,beam$^{-1}$, and the southwest ansa of the disk has a peak intensity of $1.4 \pm 0.2$\,mJy\,beam$^{-1}$. While \cite{macgregor2013} found marginally significant evidence for asymmetry in flux at the ansae of the AU Mic debris disks system, the limb asymmetry at 450\,$\mu$m is well within the uncertainties, and therefore can be considered to be insignificant. This is in line with \cite{daley2019}, which also did not find significant flux asymmetry between the ansae of the disk.

\begin{figure}[h!]
    \centering
    \includegraphics[width=8.8cm]{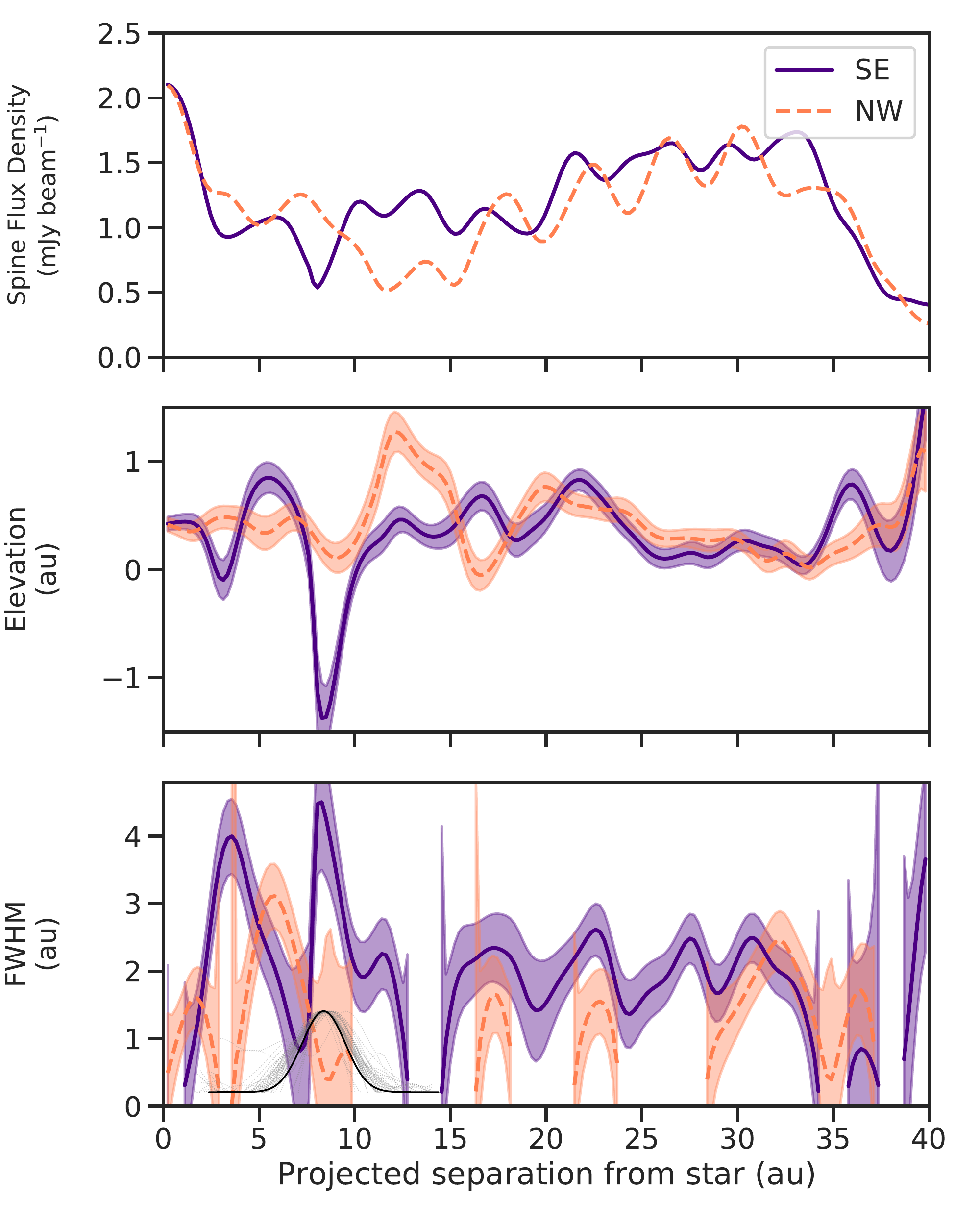}
    \caption{From top to bottom, as a function of projected separation from the star: \textbf{(a)} spine flux density, \textbf{(b)} elevation from the midplane of the disk, and \textbf{(c)} the beam-deconvolved disk FWHM. This projection confirms that we detect no significant deviations from radial and vertical symmetry. The top plot shows that, as expected, the brightness of the disk increases as a function of distance from the central star. The middle two plots confirm the results of \cite{daley2019}, in that neither warps nor fast-moving features are detected in this data. Most importantly, the bottom plot shows that the disk is marginally resolved at $\lambda=450 \ \mu$m, with a smaller scale height than measured by \cite{daley2019} at $\lambda = 1.3$ mm. Missing information in the bottom plot indicates that the disk is unresolved at those separations from the star.} 
    \label{fig:boccaletti}
\end{figure}

Before fitting any models to the disk, we can begin to characterize the vertical structure of the AU Mic disk. Figure \ref{fig:boccaletti} plots several metrics related to the disk structure. We first define a disk midplane from the position of the star at a position angle of 128.48$\degree$ (see \S 4). We then fit Gaussians perpendicular to the disk midplane, as a function of offset along the disk midplane. The top plot shows the spine flux density, which is the peak flux density of the Gaussian at each offset along the major axis of the disk. The middle plot shows the elevation, or distance of the disk's spine from the disk midplane. The bottom plot shows the beam-deconvolved full-width at half-maximum (FWHM) of each Gaussian offset along the major axis. The concept of these graphs is taken from \cite{boccaletti2015}, who plotted both elevation from the midplane of the disk and intensity of the disk as a function of projected separation from the star in order to quantify fast-moving features within the AU Mic debris disk. 

Like \cite{daley2019}, we do not find any features at the location of the fast-moving features detected by \cite{boccaletti2015}. Furthermore, the bulk of the disk mass appears to reside within a single plane, and there do not appear to be any significant warps or deviations from this midplane. The disk limbs show no significant asymmetries at this wavelength.  Most importantly, the vertical structure of the disk is resolved. The beam-subtracted FWHM is smaller at 450 $\mu$m than at 1.3 mm \citep[see][]{daley2019}, which suggests that the scale height at 450 $\mu$m might be lower than at 1.3 mm. In Section \ref{chap4} we simultaneously model the high-resolution 1.3\,mm data (from ALMA project 2012.1.00198.S, PI: A.~M.~Hughes) and 450 $\mu$m data, including the ratio of the scale height at the two frequencies.

\section{Analysis}\label{chap4}

\subsection{Modeling formalism}

To derive a best-fit model and explore the parameter space for posterior probabilities, we employed the Python package \texttt{emcee} \citep{foremanmackey2013} which implements the affine-invariant MCMC (Monte Carlo Markov Chain) algorithm described in \cite{goodman2010}. We used a Gaussian log-likelihood function ($\ln$ $\zeta =$ $ - \chi^2 / 2$) to compare a given model disk with the data. The model debris disks were created using the modeling and raytracing code described in \cite{flaherty2015}. Here we briefly recap the functional form and assumptions of the model.

The dust temperature is derived by setting the flux received by the disk from its host star equal to the flux emitted by the disk and assuming zero-albedo dust grains, such that the system is in blackbody equilibrium:
\begin{equation}
    T_{\rm dust}(r) = \left(\frac{L_{\star}}{16 \pi \sigma r^2}\right)^{\frac{1}{4}}
\label{eq:dust temp}
\end{equation}
where $L_{\star}$ is the bolometric luminosity of the system's host star, $\sigma$ is the Stefan-Boltzmann constant, and $r$ is distance from the star. The radial surface density across the disk between the inner radius of the disk ($R_{\rm in}$) and the outer radius ($R_{\rm out}$) is described by a power law:
\begin{equation}
    \Sigma (r) = \Sigma_c r^{\gamma}
\end{equation}
where
\begin{equation}
    \Sigma_c = \frac{M_{\rm disk} (\gamma + 2)}{2\pi(R_{\rm out}^{\gamma+2} - R_{\rm in}^{\gamma+2})}
\end{equation}
$\gamma$ is the power law index which describes the change in surface density with radius, and $\Sigma_c$ is the surface density normalization constant.

The vertical dust density distribution is modelled as a Gaussian function whose standard deviation characterizes the vertical structure of the disk. Recall that the aspect ratio of the disk $h$ is defined as
\begin{equation}
    h = \frac{H(r)}{r}
\end{equation}
where $H(r)$ is the scale height. Combining the previous equation with our equation for the scale height of the disk yields the dust volume density as a function of distance from the star $r$ and height above the disk midplane $z$:
\begin{equation}
    \rho_D(r,z) = \frac{\Sigma(r) \ \exp \left[-\frac{z}{H(r)}  \right]^2}{\sqrt{\pi} H(r)} 
\end{equation}
We also model the dust opacity as a power law \citep[][]{beckwith1990}:
\begin{equation}
    \kappa_\nu = 10\left(\frac{\nu}{\rm{10^{12}} \ Hz}\right)^\beta \ \rm{cm^2 \ g^{-1}}
    \label{eq:opacity}
\end{equation}
and by combining the last two equations we can describe a dust absorption coefficient $K_\nu$ as
\begin{equation}
    K_\nu = \kappa_\nu \rho_D;
\end{equation}
integrating $K_\nu$ along the line of sight from each point in the disk to Earth yields an expression for the optical depth of the disk $\tau_\nu$. We note here that the dust opacity is not trivial for this project; please see \S \ref{chap5} for further discussion.

Our final run consisted of 50 walkers and 4000 steps. We followed up our run with an autocorrelation analysis, estimating an autocorrelation length of 200 steps. Though the autocorrelation time was still rising by the end of each chain, after 10 autocorrelation lengths all parameters had levelled off such that the fractional error in the mean was no more than a few percent, and the standard error of the mean was stable; we thus removed the first 2000 steps from our run. 

\begin{table*}[h!]
    \centering
    \resizebox{\textwidth}{!}{
    \begin{tabular}{|c|c|c|c|c|}
        \hline
         Parameters & Description & Median & Best-fit & Priors \\
        \hline
         $\log_{10} M_{\rm disk}$ [M$_{\rm \odot}$] & Disk dust mass & -8.09$^{+0.02}_{-0.02}$ & -8.10 & (-10, 3) \\
         $\gamma$ & Surface density index & 2.2$^{+0.2}_{-0.7}$ & 2.1 & (-5, 5)\\
         $R_{\rm in}$ [au] & Disk inner radius & 22.1$^{+1.0}_{-0.4}$ & 21.8 & (0, 1000) \\
         $R_{\rm out, \ 1300}$ [au] &  Outer radius @ 1.3 mm & 42.0$^{+0.7}_{-0.2}$ & 42.0 & (0, 1000) \\
         $R_{\rm out, \ 450}$ [au] & Outer radius @ 450 $\mu$m & 38.3$^{+0.8}_{-0.3}$ & 38.0 & (0, 1000) \\
         $i$ [$\degree$] & Disk inclination & 88.51$^{+0.44}_{-0.05}$ & 88.51 & (0, 90) \\
         $PA$ [$\degree$] & Disk position angle & 128.48$^{+0.05}_{-0.02}$ & 128.48 & (0, 180) \\
         $f_{1300, \ \rm Mar}$ [$\mu$Jy] & Central flux density @ 1.3 mm on March 2017 & 376$^{+7}_{-6}$ & 375 & (0, 1 Jy)\\
         $f_{1300, \ \rm Aug}$ [$\mu$Jy] & Central flux density @ 1.3 mm on August 2017 & 171$^{+20}_{-25}$ & 132 & (0, 1 Jy) \\ 
         $f_{1300, \ \rm Jun}$ [$\mu$Jy] & Central flux density @ 1.3 mm, June 2017 & 222$^{+14}_{-11}$ & 221 & (0, 1 Jy) \\ 
         $f_{450}$ [mJy] & Central flux density @ 450 $\mu$m & 1.25$^{+0.08}_{-0.11}$ & 1.27 & (0, 1 Jy) \\
         $h_{1300}$ & Aspect ratio @ 1.3 mm & 0.025$^{+0.008}_{-0.002}$ & 0.023 & (0, 5) \\
         $h_{1300} / h_{450}$ & Scale height ratio @ 1.3 mm and 450 $\mu$m & 1.35$^{+0.09}_{-0.09}$ & 1.34 & (0, 4) \\
         $\beta$ & Dust opacity index & 0.06$^{+0.03}_{-0.03}$ & 0.06 & (-5, 5) \\
         \hline
    \end{tabular}}
    \caption{Results of the MCMC model fitting to both the 450 $\mu$m (this work) and 1.3 mm \citep{daley2019} data of AU Microscopii. 
    We note that ''central flux density`` refers to the flux density of the central point source (i.e. the star) in the dataset.}
    \label{tab:dmr}
\end{table*}

We initialized our MCMC run with fourteen free parameters: dust mass of the disk ($M_{\rm disk}$), inner radius ($R_{\rm in}$), disk width at both 1.3 mm and 450 $\mu$m ($R_{\rm{out}, 1300}$ and $R_{\rm{out}, 450}$), inclination ($i$), position angle (PA), surface density power law exponent ($\gamma$), flux density of the central point source for the 450 $\mu$m data ($f_{450}$) and for each of the observing sessions at 1.3 mm ($f_{1300, \ \rm{Mar}}$, $f_{1300, \ \rm{Aug}}$, $f_{1300, \ \rm{Jun}}$), the aspect ratio of the disk at 1.3 mm ($h_{1300}$), the ratio of scale heights at 1.3 mm and 450 $\mu$m ($h_{1300} / h_{450})$, and the dust opacity power law index ($\beta$).

We initially fit the 450 $\mu$m data alone. However, we found that the MCMC chains diverged into two families of solutions with different inclinations. One family had a scale height that matched the results at 1.3 mm, but a significantly lower inclination, which was not commensurate with what we know about the viewing geometry. The other family of solutions, which included the global best-fit, had a comparable inclination to the 1.3 mm results, but a lower scale height. We realized that fitting the two wavelengths simultaneously was the best way to study the relative scale heights while constraining the inclination to be the same at both wavelengths. The viewing geometry and dust properties within the AU Mic debris disk ($M_{\rm disk}$, $PA$, $i$, $\beta$) are therefore modeled to be identical at both 1.3 mm and 450 $\mu$m, while the radial and vertical structure ($h_{1300}$, $h_{450}$, $R_{\rm{out}, 1300}$ $R_{\rm{out}, 450}$) are allowed to vary at each wavelength.

\begin{figure}[h!]
    \centering
    \hspace*{-1.1cm}
    \includegraphics[width=9.5cm]{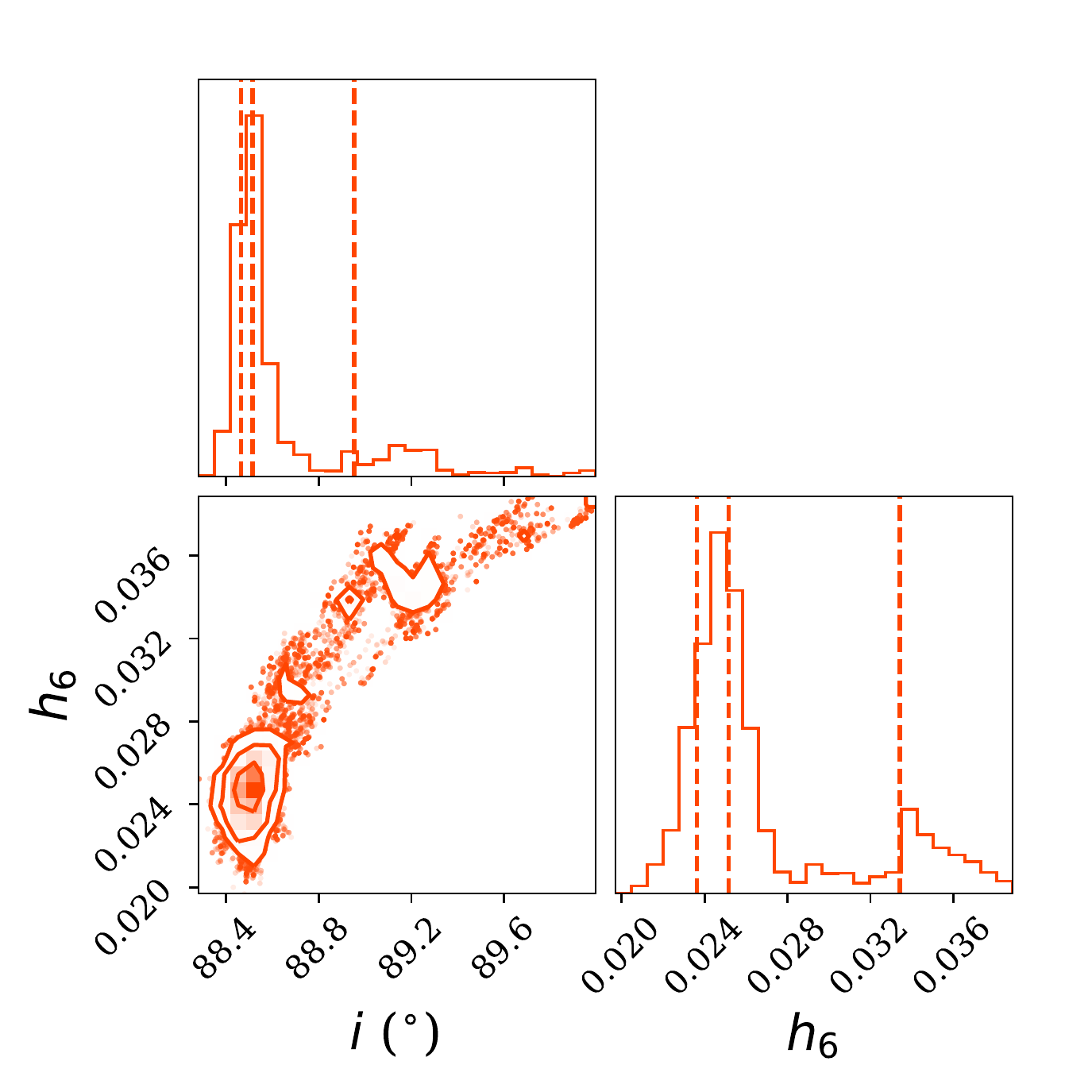}
    \caption{A visualization of the degeneracy between the AU Mic debris disk's aspect ratio at Band 6 (top histogram) and inclination (right histogram). As the modeled inclination increases, so does the aspect ratio of the disk. This plot was made using the Python package \texttt{corner} \citep[][]{foremanmackey2016}.}
    \label{fig:corner_plot}
\end{figure}

\begin{figure*}[t!]
    \centering
    \hspace*{-3cm}
        \includegraphics[width=24cm]{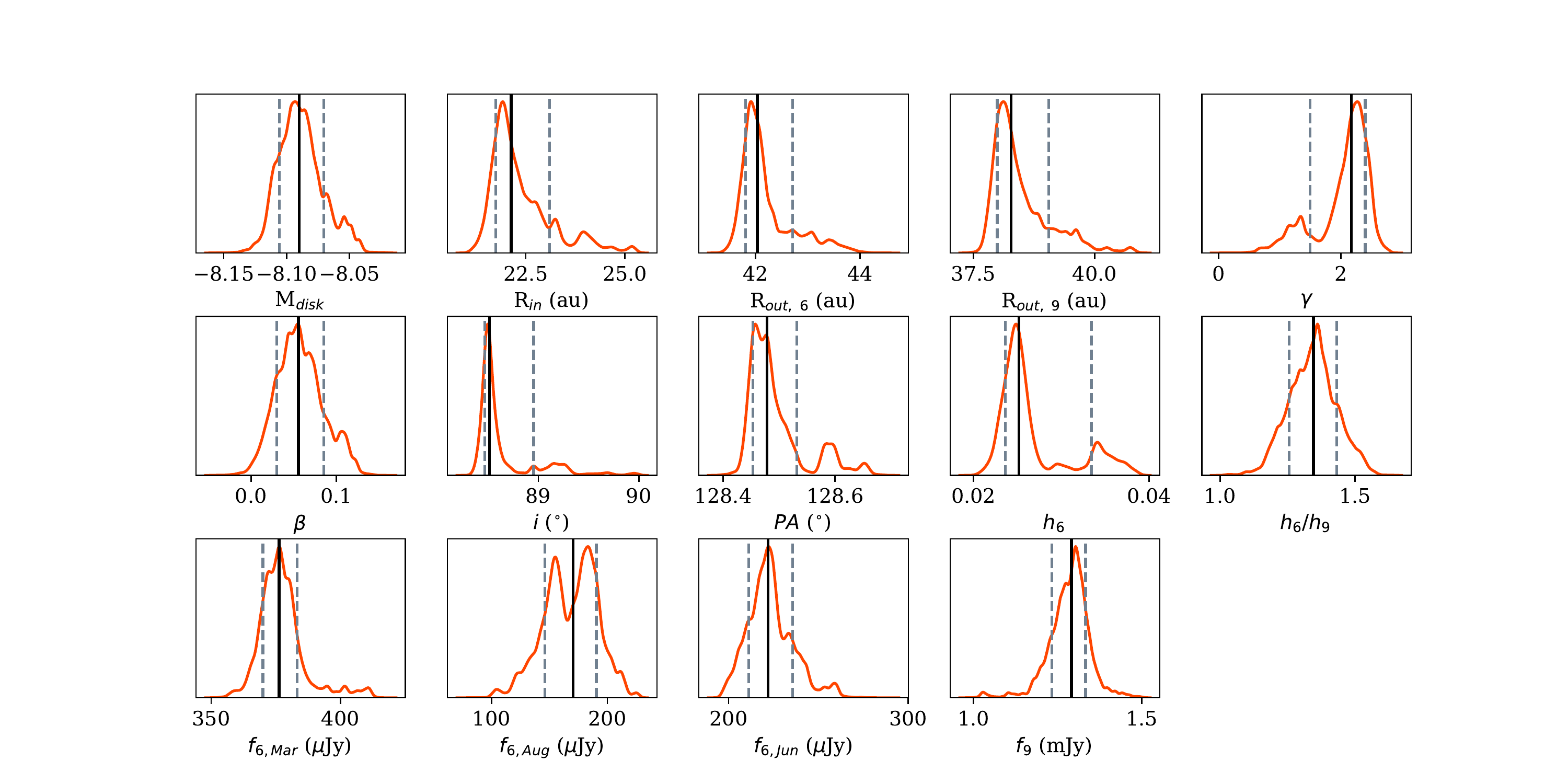}
    \caption{Kernel density estimates of the posterior probability distributions for the MCMC modeling run. The central line represents the median, and the outer dashed lines represent 16th and 84th percentiles. }
    \label{fig:histograms}
\end{figure*}

All of the disk parameters were sampled linearly except for $M_{\rm disk}$, where we sampled $\log$ $M_{\rm disk}$ instead. We also bounded the parameters with priors; for instance, we place upper bounds on the angular parameters $i$ and $PA$ of $90\degree$ and $180\degree$ respectively. We note that the lopsided posteriors in Figure \ref{fig:histograms} for both $i$ and $PA$ towards the upper bounds are not reflections of the prior boundary and are physical; please see \S 4.1 in \citet{daley2019} for further discussion of this point, including an investigation showing that relaxing the boundary on the prior and allowing it to reflect across the boundary does not change the posterior distribution for these parameters.

\begin{figure*}[t!]
    \centering
    \hspace*{-2.3cm}
    \includegraphics[width=22cm]{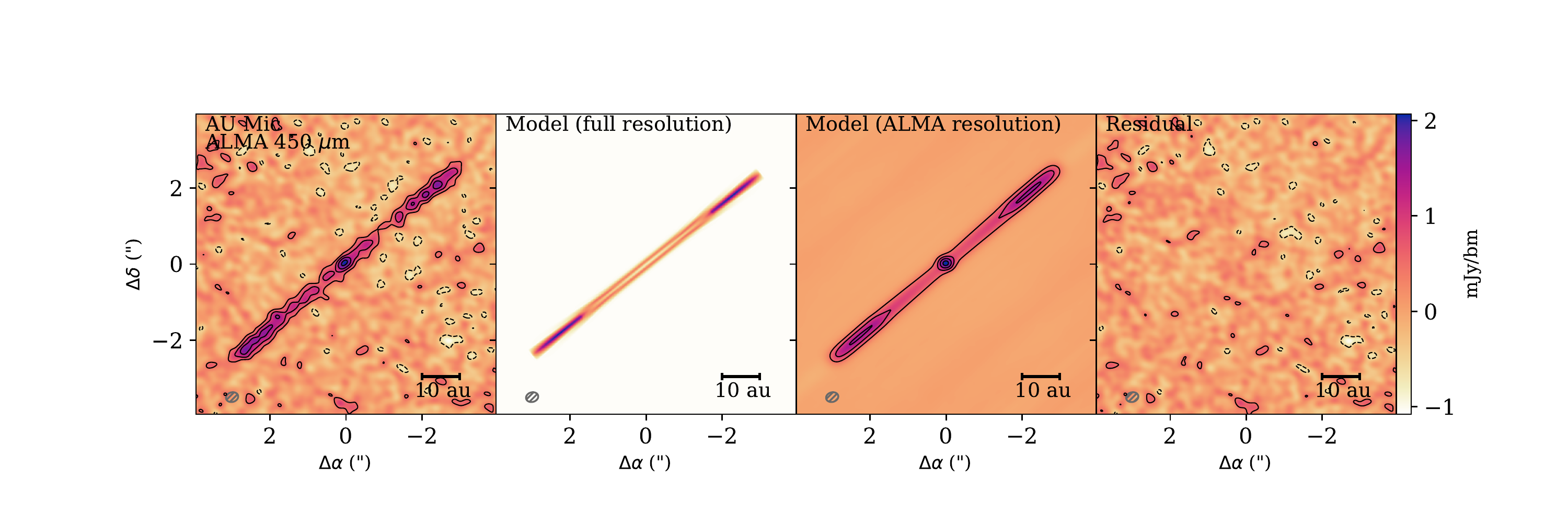}
    \caption{From left to right: \textbf{(1)} 450 $\mu$m ALMA observations of the AU Mic system. RMS noise is 240 $\mu$Jy\,beam\,$^{-1}$. The restoring beam has the same dimensions as that of Figure \ref{fig:AUMic} and is represented by the hatched ellipse in the bottom left corner. The contours represent integer multiples of 2$\sigma$. \textbf{(2)} Full-resolution image of the modelled debris disk, with the central flux set to zero. \textbf{(3)} Sky-projected image of the best-fit model disk after sampling with the same baseline lengths and orientations as the ALMA observations. The aspect ratio of the model disk is $h = 0.019^{+0.006}_{-0.001}$. \textbf{(4)} The residual generated by subtracting the model disk from the data in the visibility domain. No significant residuals exist along the major axis of the debris disk, indicating good agreement between data and model.}
    \label{fig:dmr}
\end{figure*}

\begin{figure*}[t!]
    \centering
    \hspace*{-2.3cm}
    \includegraphics[width=22cm]{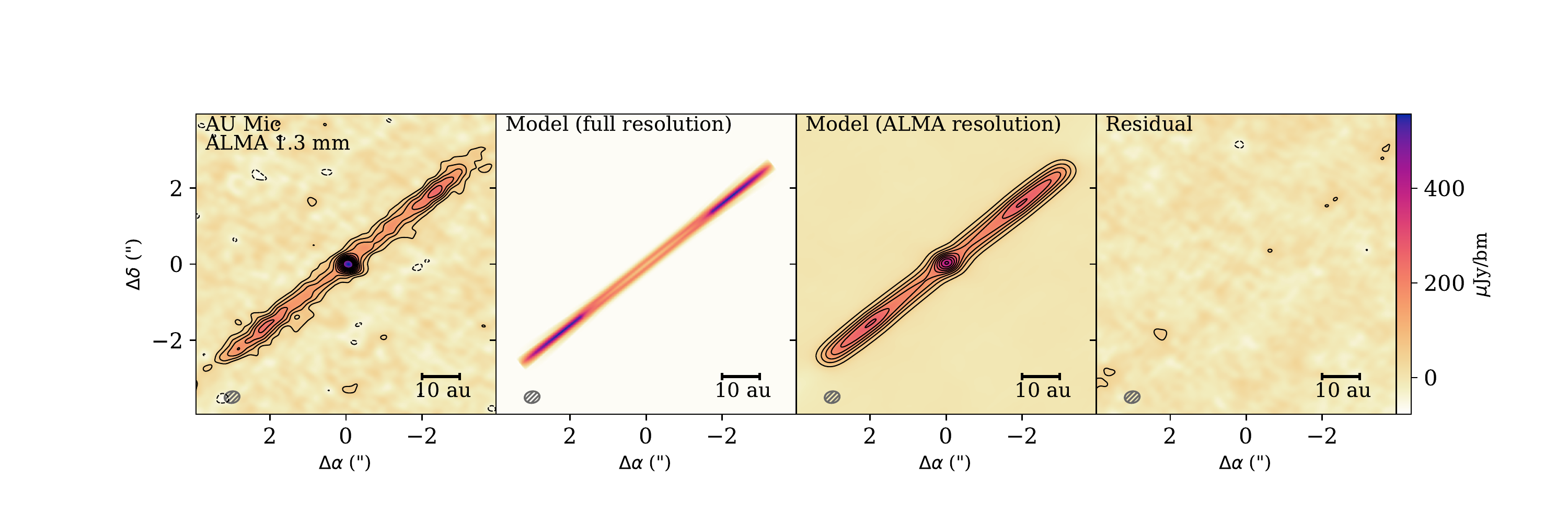}
    \caption{From left to right: \textbf{(1)} 1.3 mm ALMA observations of the AU Mic system, from \cite{daley2019}. RMS noise is 15 $\mu$Jy\,beam\,$^{-1}$, and the contours represent integer multiples of 3$\sigma$. \textbf{(2)} Full-resolution image of the modelled debris disk, with the central flux set to zero. \textbf{(3)} Sky-projected image of the best-fit model disk after sampling with the same baseline lengths and orientations as the ALMA observations. Compared to the model shown in Figure \ref{fig:dmr}, this one has an aspect ratio $h = 0.025^{+0.008}_{-0.002}$ and a different central flux due to stellar variability during observations (see \S \ref{chap3}). Since both models share the same inclination, it is evident that the scale height at this wavelength must be slightly larger than at 450 $\mu$m. \textbf{(4)} The residual generated by subtracting the model disk from the data in the visibility domain. No significant residuals exist along the major axis of the debris disk.}
    \label{fig:dmr6}
\end{figure*}

Other parameters were initialized at non-zero values, such as the aspect ratio of the disk. The best-fit parameters are obtained from the model disk with the lowest corresponding $\chi^2$ value. The model corresponding to the best-fit parameters is compared with the data in Figures \ref{fig:dmr} and \ref{fig:dmr6}, while the best-fit parameters for the AU Mic debris disk are presented in Table \ref{tab:dmr}.

A primary beam correction was applied to the model disk before using \texttt{galario} to compare the model disk to the data visibilities. Because the 450 $\mu$m data are mosaicked, three primary beam corrections are applied to the model, each corresponding to the respective location of the mosaic field. The 1.3 mm data are not mosaicked and therefore only require a single primary beam correction applied to the phase center for each execution block. After creating the model disk image, we used \texttt{galario} to convert it into model visibilities sampled at the same baseline lengths and orientations as the data, which were then subtracted from the data visibilities and Fourier transformed to yield residual images.  We found during the modeling process that the results were somewhat sensitive to the assumed size of the primary beam, due to the contrast in source vs. primary beam size at the two wavelengths.  Our investigation revealed that the primary beam size would need to deviate by $\sim$10-15\% from the value given in the ALMA technical handbook to significantly alter the results; deviations at the level of 3\% from the nominal FWHM did not produce any significant differences in the posterior distributions for the outer radius or scale height ratio. We were not able to determine an uncertainty on the primary beam size either from the technical handbook or from a helpdesk ticket.

The data, best-fit model, and residual at $\lambda = 450$ $\mu$m are shown in Figure \ref{fig:dmr}, and those at $\lambda = 1.3$ mm are shown in Figure \ref{fig:dmr6}. At both wavelengths, the best-fit model subtracts cleanly from the data in the visibility domain such that no significant residuals exist in the image (and in fact, the noise appears lower in the residual images than in the data images due to imaging artifacts caused by the CLEAN algorithm, which does not tend to work well on long, narrow flux distributions). From the data, we retrieve a scale height ratio $h_{1300} / h_{450} = 1.35^{+0.09}_{-0.09}$, and and an aspect ratio of $h_{1300} = 0.025^{+0.008}_{-0.002}$ at 1.3 mm; from the posterior distributions of these two parameters, we thus retrieve an aspect ratio of $h_{450} = 0.019^{+0.006}_{-0.001}$ for the data at 450 $\mu$m.

We remark here that for an edge-on ($i = 90\degree$) disk, there exists a natural degeneracy between inclination and scale height. To create a model that best resembles the data, the algorithm navigates a fine line between increasing the scale height (i.e. puffiness) of the disk and decreasing the inclination. However, increasing the scale height will thicken the disk at the ansae, while increasing the inclination mostly thickens the disk along the minor axis. This behavior has been described in the literature \citep[e.g.][]{graham2007,daley2019}, but is mitigated somewhat by combining the two bands and requiring that they have the same inclination. The Python package \texttt{corner} \citep[][]{foremanmackey2016} is used to visualize degeneracies within the parameter space. Like \cite{daley2019}, we find degeneracy between the scale height and inclination of the disk; this degeneracy is presented in Figure \ref{fig:corner_plot}. Nevertheless, a scale height ratio of 1 is excluded at the $3.9\sigma$ level in the posterior distribution.

\subsection{The collisional cascade indices}

The measurements made in the previous two sections allow us to characterize the collisional cascade of the AU Mic debris disk within the framework originally presented by \cite{dohnanyi1969} and modified by \cite{pan2012} to incorporate grain-size dependent velocity dispersions. The interparticle relative velocity \citep{wyatt2002} is proportional to the aspect ratio $h$, i.e.
\begin{equation}
    \langle v_{\rm rel} \rangle \approx v_{\rm Kep}(r) \sqrt{\bar{i}^2 + 1.25 \bar{e}^2} \approx 2.355 \sqrt{3} v_{\rm Kep}(r) h
    \label{eq:v_rel}
\end{equation}

Furthermore, the grain size is proportional to wavelength of observation via Equation \ref{eq:p}. The most efficient emission occurs when $\lambda = 2 \pi a$ (see \S~\ref{sec:assumptions} for a discussion of this assumption). We therefore can solve for $p$ by setting Equation \ref{eq:v_rel} equal to Equation \ref{eq:p} and taking the proportion of aspect ratios at two separate wavelengths. This yields the relationship
 \begin{equation}
     \frac{h}{h_0} = \left(\frac{\lambda}{\lambda_0}\right)^{p}
 \label{eq:scale height}
 \end{equation}
where the aspect ratio measurement $h$ corresponds to wavelength $\lambda$, while the aspect ratio measurement $h_0$ corresponds to wavelength $\lambda_0$.  The exponent $p$ is then the power law index that describes the velocity dispersion as a function of grain size. Now that the aspect ratios at both wavelengths have been modelled in the previous section, we assume that the grain size is equal to the wavelength of observation and calculate $p$ via the equation
\begin{equation}
    \frac{h_{1300}}{h_{450}} = 1.35^{+0.09}_{-0.09} = \left(\frac{1300 \ \rm{\mu m}}{450 \ \rm{\mu m}}\right)^p.
\end{equation}
Rearranging this equation to solve for the velocity dispersion index yields a value for $p = 0.28^{+0.06}_{-0.06}$. To derive $q$ for the disk, we rearrange Equation 13 from \cite{draine2006},
\begin{equation}
    \beta = (q-3) \beta_s
\end{equation}
to get
\begin{equation}
    q = 3 + \frac{\beta}{\beta_s}.
\end{equation}
where $\beta_s$ is the absorption cross-section power-law index. Using our measured value of $\beta =$ 0.06 $\pm$ 0.03 and assuming $\beta_s =$ 1.8 $\pm$ 0.2 \citep{draine2006}, we determine a value of $q = 3.03 \pm 0.02$. Because this measurement is taken from forward modeling, and not the image domain (where the line of sight for an edge-on disk passes through many pixels with different radii and therefore temperatures), this value is more reliable because it accounts for local deviations in temperature and therefore in $\alpha_{\rm PL}$. The power law indices calculated for the AU Mic debris disk, $p = 0.28 \pm 0.06$ and $q=3.03 \pm 0.02$, are the first model-dependent constraints placed on the internal strengths of colliding bodies in a disk beyond the Kuiper Belt. 

\section{Discussion}\label{chap5}

\subsection {Comparison of AU Mic disk structure with values from the literature}

In this section we compare our results to other measurements and models of the AU Mic disk.

\subsubsection{Viewing geometry}
When fitting the disk structure to both the 1.3\,mm and 450\,$\mu$m data, we assumed a common viewing geometry for both wavelengths. The measured position angle of 128.48$^{+0.05}_{-0.02}$ degrees is in strong agreement with recent measurements of the disk's position angle; for example, Cycle 0 observations of the disk with ALMA \citep{macgregor2013} found a disk position angle of 128.41 $\pm$ 0.13$\degree$, and \cite{daley2019} found a position angle of 128.50 $\pm$ 0.07$\degree$ which places our measurements well within 1$\sigma$ of their work.

The inclination of the disk is less trivial, with values in the literature reaching as high as 89.4$\degree$ \citep{krist2005} or assumed to be 89.5$\degree$ \citep{macgregor2013, schuppler2015}. Our measurement of the inclination is 88.51$^{+0.44}_{-0.05}$ degrees is in strong agreement with the inclination measurement of \cite{daley2019}, $i = 88.5^{+0.3}_{-0.2}$. Since the 450 $\mu$m data are less vertically extended than the 1.3 mm data, they provide a more stringent constraint on the disk inclination. Assuming a common geometry thus restricts the 1.3 mm data to a region of parameter space with higher $i$ and larger $h$ than in \cite{daley2019}. 

\subsubsection{Radial structure}
\label{sec:radial_structure}

The inner radius of the disk at 450 $\mu$m is 22.1$^{+1.0}_{-0.4}$ au. There is tentative evidence in the literature for the presence of an inner ring of debris \citep[e.g.][]{matthews2015}; \cite{daley2019} modelled as being at a radius of 11.9$^{+1.7}_{-1.8}$ au. If we consider the AU Mic system to be analogous to the Kuiper Belt, this inner ring of debris could be considered analogous to our Solar System's asteroid belt which lies between the orbits of Mars and Jupiter. We do not include a ring in our fit to the data, as given the lower signal-to-noise ratio (SNR) in the 450 $\mu$m data compared to the 1.3 mm data, we would not expect to detect it. The inner radius is also much farther than the two planets which orbit AU Mic at semi-major axes of $a = 0.0645 \pm 0.0013$ au \citep[][]{plavchan2020} and $1.1101 \pm 0.0022$ au \citep[][]{martioli2020} respectively. This could imply the presence of another planet in the system in order to account for the location of the inner disk, and which could potentially stir the collisional content of the disk. 

Recent work by \cite{pearce2022} found, after adopting an inner disk radius for the AU Mic system of 28.7 $\pm$ 0.3 au, that a planet with a mass of 0.44 $\pm$ 0.03 M$_{\mathrm{Jup}}$ and semi-major axis of 21.9 $\pm$ 0.3 au would be able to ``sculpt'' the inner part of the AU Mic debris disk. Since our model yields a different inner radius than in \cite{pearce2022}, we recalculated this planet mass and radius using their publicly available code\footnote{\url{https://github.com/TimDPearce/SculptingPlanet}}, finding that a planet with a minimum mass of 0.34 $\pm$ 0.03 $M_{\rm Jup}$ and maximum semimajor axis of 17.0$^{+0.7}_{-0.3}$ au would be able to truncate the inner part of the disk that is modeled in this work. 

It is not trivial to assume a common radial geometry, and we ultimately allow the model to have a different outer radius for each data set. We do not vary the inner radius with wavelength since our signal-to-noise in the 450 $\mu$m data is lower than 1.3 mm and therefore cannot provide insight into the low-SNR features that tentatively indicated an inner ring in that data set. We define the outer radius ratio to be analogous to our scale height ratio; we find that $\Delta R_{1300} / \Delta R_{450} = 1.23^{+0.02}_{-0.03}$; in other words, the outer radius of AU Mic's debris disk is over 20\% larger at 1.3 mm than at 450 $\mu$m. 

The outer radius at 1.3 mm for our model is measured to be 42.0$^{+0.7}_{-0.2}$ au, which deviates slightly above 1$\sigma$ from the best-fit of 41.5$^{+0.4}_{-0.5}$\,au measured by \cite{daley2019} and by just under 3$\sigma$ from the outer radius of $40.3 \pm 0.4$\,au measured by \cite{macgregor2013}. Our results at 1.3 mm for the outer radius differ from previous measurements because we are able to constrain our model with the 450 $\mu$m data set, partially assuming a common geometry at both wavelengths. The outer radius at 450 $\mu$m, on the other hand, is measured to be 38.3$^{+0.8}_{-0.3}$ au, which is significantly lower than at 1.3 mm. We did not attempt to constrain the shape of the outer edge at either wavelength, since the lack of residuals for the best-fit model is consistent with an abrupt cutoff.

Naively one would expect that the smaller grains would exist farther away from the star than larger grains. Several works \citep[e.g.,][]{schneider2014,ahmic2009,wilner2011,wilner2012} describe ways in which the outer radius of a debris disk can differ between millimeter and OIR wavelengths, primarily due to the ``halo" of small grains blown out by stellar winds and radiation pressure. We see the opposite however; the larger grains appear to lie farther from the star than the smaller grains. 

Another possibility to explain the difference in radii at the two wavelengths is gas drag, similar to a modified version of the effect discussed in \citet{powell2017}.  However, the direction of the gas drag effect should predict a smaller outer radius at longer wavelengths, which is the opposite of the direction we observe.  On the other hand, stellar winds moderated by giant stellar outbursts as described in \citet{chiang2017} could induce ram pressure that would make collisionally created submillimeter grains somewhat more eccentric than collisionally created millimeter grains at birth, which would predict a difference in radii in the direction that we see. 

\subsubsection{Vertical structure}
The disk appears thinner at 450\,$\mu$m than at 1.3\,mm in the image domain, and this result is confirmed via modeling in the visibility domain. Since the disk at 1.3\,mm is 35\% puffier than at 450\,$\mu$m, these results imply that the 450\,$\mu$m-sized grains in the disk have a smaller velocity dispersion than the 1.3\,mm-sized grains, and indicates that there is likely some damping of velocities within the disk. 

An aspect ratio at $\lambda=450$ $\mu$m of 0.019$^{+0.006}_{-0.001}$ at a radius of 40 au implies a scale height at 40 au of 0.76$^{+0.24}_{-0.04}$ au. The scale height represents the standard deviation of a Gaussian distribution of dust particles; it is thus possible, alternatively, to describe the vertical structure of the disk by measuring the full-width at half maximum (FWHM) of this distribution. The relationship between the FWHM and the standard deviation $\sigma$ is ${\rm FWHM} = 2 \sqrt{2 \ln 2} \sigma$, so that the FWHM of the disk at 450 $\mu$m is 1.79$^{+0.56}_{-0.09}$ au. The scale height at 40 au for the 1.3 mm dust emission is 1.00$^{+0.32}_{-0.08}$ au, and the FWHM is 2.35$^{+0.75}_{-0.19}$ au. Because our modelled scale height at 1.3 mm is slightly lower than what was modelled by \cite{daley2019}, our FWHM is also slightly lower.

It is important to note that when examining how the half-width at half-maximum (HWHM) of the disk changes as a function of wavelength, discrepancies between modelled FWHM and apparent FWHM appear across the literature for AU Mic observations due to differences in assumed or fitted inclination, and so for this comparison only models of the AU Mic debris disk will be discussed. There seems to be some evidence (see Table \ref{tab:fwhm}) that the FWHM, and hence the scale height, decreases as a function of wavelength in the OIR regime. The main outlier is \cite{schuppler2015}, who modelled the dust production and dust dynamics from collisions in the AU Mic debris disk, and attempted to fit the resulting collisional models to many different SED points from scattered light \citep[][]{fitzgerald2007, schneider2014} to $\lambda =$ 1.3 mm \citep[][]{macgregor2013}; this work was not intended for fitting images, and the approximate FWHM given in Table~\ref{tab:fwhm} results from an approximation relating their reported eccentricities to inclinations in equipartition, which may be considered an overinterpretation. Alternatively, it is possible that scale height (and ultimately the velocity dispersion of grains within the disk) may fluctuate rather than monotonically increase/decrease as a function of wavelength. More observations of the disk at other wavelengths are necessary to determine whether there is a global trend of disk scale height with wavelength.

\begin{table}
    \centering
    \resizebox{0.5\textwidth}{!}{
    \begin{tabular}{|c|c|c|c|}
    \hline
         Model reference & $\lambda$ ($\mu$m) & $i$ ($\degree$) & FWHM (au) \\
         \hline
         \citet{daley2019} & 1350 & 88.5 & 2.9$^{+0.5}_{-0.4}$ \\
         This work (2022) & 1350 & 88.51 & 2.35$^{+0.75}_{-0.19}$ \\
         This work (2022) & 450 & 88.5 & 1.79$^{+0.56}_{-0.09}$ \\
         \textit{\citet{schuppler2015}} & $>100$ & 89.5 & 0.4 \\
         \textit{\citet{schuppler2015}} & $< 70$ & 89.5 & 4 \\ 
         \citet{metchev2005} & 0.647 - 1.63 & 89 & 1.6 \\
         \citet{krist2005} & 0.430 -  0.833 & 89.4 & 1.8 \\  
         \hline
    \end{tabular}
    }
    \caption{The full-widths at half-maximum (FWHMs) of several models of the AU Mic debris disk, measured at a distance of 40 au from the star, across a wide range of wavelengths. All values for FWHMs are derived from Table 4 in \cite{daley2019}. There appears to be marginal evidence for an increase in scale height of the disk as a function of wavelength, though the measured variations in scale height with wavelength are complex. Modeling methods employed by \cite{schuppler2015} are different enough from other studies (and not intended for direct comparison with vertical structure measured from images) that they may not provide a good comparison for the FWHM of the disk.}
    \label{tab:fwhm}
\end{table}

\subsubsection{Dust mass and opacity}

The dust mass that we derive depends strongly on the assumed mass opacity. We measure a dust mass of (2.7 $\pm$ $0.1) \times 10^{-3} \rm{M}_{\earth}$ within the AU Mic debris disk. This dust mass is over three times smaller than the measurement of (9.28 $\pm$ 0.13) $\times$ $10^{-3}$ M$_{\earth}$ made by \cite{daley2019} for an opacity of 2.3 cm$^2$ g$^{-1}$. It is also smaller than the dust mass of 7 $\times$ 10$^{25}$ g (approximately 1.2 $\times$ 10$^{-2}$ M$_{\earth}$) reported by \cite{macgregor2013} by adopting an opacity of 2.7 cm$^2$ g$^{-1}$ and assuming a dust temperature at 35-45 au of 25 K, and smaller than the dust mass of $\sim$ 0.01 M$_{\earth}$ calculated by \cite{strubbe2006} by fitting a thermal spectrum to an assumed steady-state collisional cascade. 

With regard to the dust opacity, it is not entirely physical to vary the dust opacity index $\beta$ without altering the dust opacity constant $\kappa$ as well, although their dependence on one other is complex and ambiguous. As Equation \ref{eq:opacity} states, we assume a normalization constant of 0.1 cm$^2$ g$^{-1}$ for our model regardless of the value of $\beta$. However, the effect of this assumption returns a lower dust mass of the disk than previous measurements. It was necessary to allow variations in $\beta$ in order to fit the flux precisely at both wavelengths.  We left the opacity normalization unchanged at a frequency of 100\,GHz because our main focus is on the spatial distribution of flux rather than the conversion from total flux to mass (which is a separate though no less interesting problem), and allowing $\beta$ and $M_{\rm disk}$ to vary ensures that the flux will be normalized at both wavelengths.  

\subsubsection{Modeling assumptions}
\label{sec:assumptions}

There are a number of assumptions that we make in our modeling process that should be highlighted; deviations from any of these assumptions could alter the interpretation of our observational results. 
We assume that the grain size is approximately equal to the wavelength, i.e. $\lambda \approx 2 \pi a$.  This assumption rests on two other simplifying assumptions: (1) that the grain size distribution follows a smooth power law index $q$, with values comparable to a Dohnanyi distribution so that the number of particles (and therefore emitting surface area) is strongly weighted towards smaller particles, and (2) that the emission efficiency of a dust grain falls off as $1/\lambda$ for wavelengths longer than $\approx 2 \pi a$.  The net effect of these two power law distributions is that the smallest grain that can emit efficiently (namely, one with size comparable to the wavelength of observation) will dominate the emission at a given wavelength.  However, one can imagine a situation where one grain size dominates the total grain size distribution within the disk, which would yield a weaker scaling relation and hence a larger value for $p$. Alternatively, an optically thicker disk would return a lower value on $p$, since the wavelength scaling relation would become less sensitive to particle size. 

We utilize Equation 13 from \cite{draine2006} as a crude approximation of the grain size distribution at submillimeter wavelengths; to retrieve a more accurate grain size distribution index for the disk (at sub-mm and mm wavelengths), it is necessary to fully model the dust and grain composition within the disk. Recently, \cite{arnold2021} attempted to model the grain composition of the AU Mic disk at scattered light wavelengths, but found that it was possible to model the spectrum of the disk with many different compositions.

There are additional uncertainties associated with our derivation of $q$. For instance, when calculating $q$ for our disk, we assume a dust opacity power law index from \cite{draine2006} of $\beta_s = 1.8 \pm 0.2$. It is critical to note that \citet{draine2006} used Equation \ref{eq:q} to explain why protoplanetary disk opacity spectral indices $\beta$ were lower than the expected opacity spectral indices for interstellar medium $\beta_{\rm ISM} = 1.7$. However, here we are taking our measured $\beta$ and assuming a value for $\beta_{\rm ISM}$ to retrieve the grain size distribution index, effectively the inverse of the work done in \cite{draine2006}.

Furthermore, $\beta_s$ is sensitive to the composition of the debris disk. \cite{lohne2020} demonstrated a link between the millimeter spectral index $\alpha_{\rm mm}$ and the grain size distribution index $q$, which implies that the size distribution of AU Mic can vary with the assumed disk composition. By using data from \cite{macgregor2016} to calculate the Planck and millimeter spectral indices, \cite{lohne2020} showed that if AU Mic's disk was composed primarily of water ice, $q$ would be close to 3, whereas a composition of amorphous carbon or pyroxene would return a value greater than the Dohnanyi index $q = 3.5$. But for the most part, \cite{lohne2020} retrieves values for $q$ from his analysis that are larger than what would be retrieved via the cruder approximation from \cite{draine2006}.

There is also an inherent inconsistency in assuming a constant value of q throughout the disk while also measuring different scale heights for two different grain sizes (because if you have different scale heights for different grain sizes, then the grain size distribution must change with height above the midplane).  Fundamentally, we cannot tell whether the scale height is different for the two wavelengths because one size of grain is not present above a certain height (which we interpret as a difference in velocity of grains seen at two wavelengths and thus is reflected by $p$), or whether the grains are present but the composition changes in such a way as to make them fainter at one wavelength than another (which would mimic as a change in $q$).  We have a theoretical prediction that favors the former scenario (which is why we favor that explanation in our analysis), but we do not have reason {\it a priori} to expect a change in composition with scale height.  However, the current data cannot rule out either scenario.  

Finally, central assumptions of \citet{pan2012} are that the dust is in a steady state and is not subject to radial transport. Given the observed fast-moving features in scattered light, there are reasons to expect some non-steady-state processes in the AU Mic disk. As discussed in \S~\ref{sec:radial_structure}, the likelihood of stellar eruptions and the known flaring behavior of AU Mic both suggest variations in radiation and stellar wind rates that could perturb the disk. Estimates of the stellar wind mass loss rate in the AU Mic system range from $\sim5-600$ times the solar wind value \citep{augereau2006,kavanagh2021,alvarado2022}. These effects should be more pronounced on the small grains than on the large grains, but the difference in outer radii between the 450\,$\mu$m and 1.3\,mm images does suggest that stellar winds could be playing a role in shaping the millimeter morphology of the disk. In addition, a transport-dominated disk would be expected to exhibit a flatter size distribution than a typical disk \citep{reidemeister2011,wyatt2011}, which could explain why the measured value of $q$ in the AU~Mic system is unusually flat compared with other debris disks \citep{macgregor2016}.

On the other hand, radial transport would only significantly affect the interpretation of our results if the transport timescale is short compared with the collision timescale for the grains.  Millimeter grains are probably too large for Poynting-Robertson drag to be relevant, but stellar wind drag could be relevant in this size range.  We can roughly estimate the transport timescale by figuring out how long it takes for the mass of solar wind particles landing on a grain to be equal to the grain's original mass (assuming that the stellar wind is confined within about 1 Sr):
\begin{equation}
    m_\mathrm{dust} \approx \dot{M_*} \Omega_\mathrm{dust} t_\mathrm{rad}  
\end{equation}
where $m_\mathrm{dust}$ is the mass of a dust grain, $\dot{M_*}$ is the stellar wind mass loss per time, $\Omega_\mathrm{dust}$ is the solid angle subtended by a dust grain, and $t_\mathrm{rad}$ is the radial transport timescale.  This relationship yields a transport timescale of a few Myr for 1\,mm-size dust grains and a stellar wind mass loss rate of 10$\times$ the solar value, which is long compared with the expected 100,000-yr collision timescale expected at a distance of 40\,au from a 0.5\,M$_\sun$ star in a disk with a vertical optical depth of order $10^{-3}$. Since the latter timescale is for collisions between similar-size particles, it is a firm upper bound on the timescale for catastrophic collisions.  A more careful calculation conducted by \citet{strubbe2006} yields drift timescales that are even longer than this rough estimate for similar parameters.  It is also important to note that the upper end of the estimated stellar wind rates for the AU Mic system are larger by a factor of 60, which shortens the timescale to a value more comparable with the age of the system. While the transport timescale does approach the age of the system, as long as the collisional timescale is shorter than the transport timescale, the collisional cascade would still be expected to operate in the standard (non-transport-dominated) way. 

\subsection{The AU Mic Collisional Cascade}

In \S~\ref{chap4} we derive values for the velocity dispersion power law index, $p = $ 0.28 $\pm$ 0.06, and the grain size distribution power law index, $q = $ 3.03 $\pm$ 0.02, via modeling the AU Mic debris disk. Here, we describe the AU Mic collisional cascade within the context of theoretical models for collisional cascades, and to compare our results with measurements of $p$ and $q$ within the Kuiper Belt. We note here that the millimeter and sub-millimeter data employed in this paper constrain the collisional cascade to $\approx$ millimeter-sized debris particles. As we discuss shortly, these measurements could fluctuate due to variations \citep[e.g.][]{fraser2009, belyaev2011} in the overall size distribution of debris within the disk, and that our comparison to the Kuiper Belt in particular requires some extrapolation. 

\subsubsection{Comparison with collisional cascade models}

A pioneering study by \cite{dohnanyi1969} was the first to characterize how the grain size distribution would equilibrate in the context of a steady-state collisional cascade, finding analytically that the grain size distribution index $q = 3.5$, for disks under steady-state conditions and a few further simplifying assumptions. Other analytical \citep[e.g.][]{williams1994, wyatt2007} and numerical models \citep[e.g.][]{obrien2003, krivov2007} recovered the Dohnanyi solution for steady-state collisions within a debris disk. 

\cite{pan2012} extended this work to consider how the velocity dispersion would equilibrate for grains of different sizes. Similarly to the cascade equations in \S~\ref{chap1}, they parameterize the collisional cascade indices $p$ and $q$ via Equations 1 and 2. They analytically described several mechanisms that would damp the motion of bodies within the collisional cascade and hence affect the measured indices. They present a complete formalism for steady-state collisional cascades; the model presented in \cite{dohnanyi1969}, for example, is a consequence of setting $p=0$ in the strength regime (i.e. no velocity damping), and the Dohnanyi grain size distribution index $q=3.5$ results from setting the energy per unit mass $Q^* (a)$ to be constant i.e. independent of grain size $a$.

In \S~\ref{chap4} we determine values for the velocity dispersion power law index $p = 0.28 \pm 0.06$ and the grain-size distribution index $q = 3.03 \pm 0.02$ by modeling the radial and vertical density structure of the AU Mic debris disk at both 450\,$\mu$m and 1.3 mm simultaneously. These parameters place AU Mic's collisional behavior within the gravity regime per \cite{pan2012}, where the velocities of small bodies are greater than large bodies, and where catastrophic collisions significantly damp the velocities of bodies within the disk. In this regime, $0.20 < p < 0.25$ and $3.21 > q > 3$. 

Catastrophic collisions in the gravity regime have been described in theoretical work. For steady-state collisional cascades, several works \citep[][]{williams1994,tanaka1996,pan2005} retrieve the Dohnanyi grain-size distribution power law index $q=3.5$. Describing a semi-steady collisional cascade, \cite{pan2005} derive $2.87 < q < 3.14$ for a collisional population of objects whose binding energies are dominated by gravity and whose planetesimal population decreases over time. 

Some works have attempted to model the conditions necessary for catastrophic collisions. \cite{leinhardt2009} conducted full numerical simulations of catastrophic collisions and found that shear strength is important in determining collisional outcomes for km-sized objects within the gravity regime. \cite{leinhardt2012} used numerical methods to explore collisional outcomes, including catastrophic and super-catastrophic collisions, between bodies in the gravity regime; however, this work mostly focused on collisional outcomes in the context of planet formation. Other works have attempted to model collisions in the context of debris disks; \cite{gaspar2012} numerically modelled collisional cascades in debris disks and found a steeper $q = 3.65 \pm 0.05$ when accounting for both erosive and catastrophic collisions within the disk and when considering conditions such as varying tensile strength of object material; this value is slightly steeper than the Dohnanyi solution retrieved by other analytical works in the literature. Future theoretical work could build upon \cite{pan2012} by accounting for the effects of grain composition on collisional outcomes and on the equilibrium conditions of the collisional cascade.

Supporting evidence for catastrophic collisions in the AU Mic debris disk is not difficult to find. The fast-moving features detected by \cite{boccaletti2015} and later confirmed by \cite{boccaletti2018} have been described as “dust avalanches” \citep{chiang2017} where sub-$\mu$m dust particles gain mass as they shatter other planetesimals in the disk; these outflows are attributed to the stellar winds. Our observations did not detect these features at 450 $\mu$m; however, the same destructive collisions which yield ``dust avalanches'' may also be responsible for the damping of velocities within the disk. 

It is certainly surprising to place observations nominally probing millimeter grains within the gravity regime.  Previous work has found that the strength-gravity regime boundary lies at sizes of order hundreds of meters \citep{benz1999}, or perhaps as small as a few meters for extreme models like loosely bound pebble piles \citep{krivov2018}, so we would expect millimeter grains to lie in the strength regime instead.  Ultimately, our measurements of $p$ and $q$ are probing the power law $\gamma$ that is used to describe the power law index of the critical fragmentation energy, $Q^* \propto a^\gamma$.  Using the relationship $\gamma = \frac{21 - 2p - 6q + 2qp}{q-1}$ \citep{kobayashi2010,pan2012}, our measured values of $p$ and $q$ imply a value of $\gamma = 1.95$.  Fundamentally, this means that it takes less energy to destroy a small grain than a larger one.  While this property is normally associated with the gravity regime, it doesn't necessarily require that a grain's self-gravity is the physical reason for this, but it is not immediately obvious what else could explain these results.

One alternative possibility is that the millimeter grains are actually in the strength regime, but that the size distribution is not a simple power law, and instead exhibits ``ripples'' or ``waves" as a function of grain size \citep[e.g.,][]{campo-bagatin1994,wyatt2011}.  Such ripples occur in response to a sharp cutoff in the grain size distribution, for example due to radiation pressure or stellar wind pressure that preferentially removes grains smaller than a certain size. The relevance of this effect depends on the ``wavelength" of the waves, which equals the size ratio between a body and the smallest particle that can collisionally destroy it.  In order for the effect to strongly alter the interpretation of the observations, the wavelength would need to be comparable to (i.e., not much shorter or longer than) the ratio of the two wavelengths we used to measure the change in brightness and vertical structure -- in this case, $\sim3$.  As a rough estimate, a scale height of a few percent at a distance of 40\,au from a 0.5\,M$_\sun$ star corresponds to a collisional velocity of order 100\,m\,s$^{-1}$, so that the energy per unit mass delivered by an incoming projectile in a collision would be $10^8$\,erg\,g$^{-1}$.  If the energy per unit mass needed to destroy a grain is $Q^*$, and all the same size grains have the same mass density, then for a catastrophic collision we require:
\begin{equation}
\frac{\mathrm{target~size}}{\mathrm{projectile~size}} \approx \left( \frac{10^8\,\mathrm{erg\,g^{-1}}}{Q^*}\right)^{1/3} \ .
\end{equation}
This ratio is the ``wavelength" of the waves.  Estimates for $Q^*$ vary widely, up to $10^7$\,erg\,g$^{-1}$ for small ($<1$\,km) sizes \citep[e.g.,][]{holsapple1993,benz1999}, which suggests that the wavelength could conceivably be comparable to the factor of $\sim$3 between the band frequencies.

This effect also depends on the amplitude of the waves, however. The amplitude might be particularly low for AU Mic compared to other systems, because a sharp cutoff in the size distribution due to radiation pressure is not expected for this low-luminosity star. AU Mic is so red that the dust grain size at which the stellar luminosity $L_*/c$ balances gravity is small compared to the peak wavelength of the stellar blackbody, which means that absorption of AU Mic's radiation by these grains will be inefficient and they will not become unbound.  Absorption at smaller dust grain sizes is even less efficient, so no sharp cutoff in the grain size distribution is expected due to radiation pressure. However, if a stellar wind operates that is strong enough to cause the transport time to dominate over the collision time, a sharp grain size cutoff could be introduced.

\subsubsection{Comparison with the Kuiper Belt}

The only other debris disk with a comparably characterized collisional cascade is the Kuiper Belt. While observations support a rich collisional history in the Kuiper Belt, a definitive model of the collisional evolution of Kuiper Belt objects remains an open question \citep{kenyon2004, leinhardt2007}. \cite{pan2005} applied their work to the Kuiper Belt and determined that Kuiper Belt objects can be described as ``strengthless'' rubble piles which are prone to shattering. 

Within the Kuiper Belt, the indices $p$ and $q$ that describe the collisional cascade have been measured across many different size regimes. The velocity dispersion of Kuiper Belt objects (KBOs) of all observed sizes is about 1\,km\,s$^{-1}$ \citep[][]{leinhardt2008}, which suggests a value for $p = 0$, though \cite{abedin2021} showed that non-negligible differences in collisional velocities exist between different populations of Kuiper Belt objects. It is more complicated to describe the dust grain size distribution of the Kuiper Belt for two reasons. Astronomers are mainly able to observe the km-sized objects in the belt as opposed to the dust, since the lower mass of dust in the Kuiper Belt results in a lower collision rate, so that most of the dust is ejected by the large planets out of the Solar System, and the rest spirals in towards the star due to Poynting-Robertson drag \citep{yamamoto1998}, whereas in debris disks around distant stars the km-size objects are invisible and only the dust can be observed. Furthermore, the grain size distribution power law slope varies dramatically as a function of grain size in the Kuiper Belt, unlike the smooth power law assumed for debris disks. Nevertheless, a Dohnanyi grain-size distribution power law index of $q=3.5$, and more generally the power law relationship $\frac{dN}{da} \propto a^{-q}$, has been shown to agree broadly with observations of Kuiper Belt objects \citep[e.g.][]{bernstein2004,kenyon2004, schlichting2013}.

It is also very difficult to observe Kuiper Belt dust due to foreground emission from the zodiacal cloud, and thus dust is typically modelled \citep[e.g.][]{poppe2016}. Within the mm/sub-mm regime, the modeled grain size distribution index has varied in different works between $q \sim 2.5$ \citep{moromartin2003} to $q < 3.5$ \citep[][]{vitense2012}, to $q = 3.5$ \citep[][]{kuchner2010}, with the caveat that there are likely waves in the size distribution so that the measured slope $q$ is sensitive to the particle sizes included in the measurement. The dust populations that can be probed lie within the $\mu$m regime, and thus are difficult to compare with the mm-sized dust in this work; these populations can be constrained today using the Venetia Burney Student Dust Counter (SDC) on the New Horizons spacecraft.

\cite{poppe2019} combined modelling with dust measurements from the SDC to constrain the dust grain production in the Kuiper Belt, finding that the Kuiper Belt emits the most thermal emission within the 50-100\,$\mu$m regime, with sculpting of the belt's inner radius by Jupiter clearly detectable. This sculpting feature is also apparent at 500\,$\mu$m, though the Kuiper Belt at this wavelength is less bright; nevertheless, perhaps this is a supporting argument for the existence of a planet shaping the inner radius of the AU Mic disk \citep[as argued by][]{pearce2022}. \cite{bernardoni2022} further studied the $\mu$m-sized dust (0.356 $<$ $r_g$ $<$ 2.370 $\mu$m) using detections from the SDC, finding that the overall flux estimates of the Kuiper Belt agree with predictions from modelling \citep[][]{poppe2016,poppe2019}. However, neither of these works calculate the size distribution index for $\mu$m-sized dust in the Kuiper Belt, as has been done for km-sized objects.

We emphasize that the value for $q$ measured for AU Mic's debris disk in this work purely describes mm to sub-mm grain populations. It is possible that, as has been modelled for the Kuiper Belt \citep[][]{kuchner2010}, the $p$ and $q$ values we measure for the disk are sensitive to waves in the overall distribution of velocity dispersion and grain size  within the AU Mic disk. The only way to verify this hypothesis is via observations of the disk at many different wavelengths. Also noteworthy is work by \cite{arnold2021}, who modelled three different shapes of sub-micron ($\approx$ 0.2 $\mu$m) dust grains in the AU Mic debris disk and concluded that the value of $q$ measured by fitting the scattered light spectrum varied dramatically from $q = 3.72^{+0.43}_{-0.28}$ to $q = 4.57^{+0.28}_{-0.40}$ \citep{lomax2018}. To summarize, the values of $p$ and $q$ that we measure in the AU Mic debris disk are broadly comparable to those measured for the Kuiper Belt and consistent with the range predicted by the physics of the collisional cascade, although the comparison is complicated by our inability to directly compare values for the same range of grain sizes and by our lack of knowledge of the composition of the dust grains.  

\subsection{Future work}

Evidently, variations must exist in the overall grain size distribution for the AU~Mic system; in order to draw more general conclusions about AU Mic's collisional cascade across wider ranges of grain sizes, it is necessary to make high-resolution, multi-wavelength observations of the vertical structure of the disk within other portions of the electromagnetic spectrum. While \cite{fitzgerald2007} performed multi-wavelength observations of the disk in the infrared, higher resolutions are necessary to determine whether the trend of increasing scale height continues for $\mu$m and sub-$\mu$m grains. Probing the cascade behavior of smaller or larger grains via observations at other wavelengths will make it possible to determine to what extent AU~Mic's debris disk can be described by a Dohnanyi size distribution power law (i.e. $q = 3.5$) and to what extent AU~Mic is consistent with theoretical predictions for a steady-state collisional cascade. With regards to AU Mic, Band 3 -- or, eventually, Band 1 -- observations of the disk would be able to extend our cascade description further into millimeter wavelengths. It should also be possible to use the Square Kilometer Array (SKA) or, eventually, the Next Generation Very Large Array (ngVLA) to observe at even longer wavelengths. These observations would allow us to determine both whether waves exist within the collisional cascade (i.e. the size distribution and velocity dispersion indices) at millimeter wavelengths, as well as whether the trend of increasing FWHM extends beyond into the centimeter-wavelength regime. 

Some testable predictions include that if the disk is in fact transport-dominated at the wavelengths probed by our data, we would expect to see a steepening of the power law index $q$ at longer wavelengths, because even in the AU~Mic system those larger grains would no longer be transport-dominated. Likewise, for disks around earlier-type stars with lower stellar wind rates, we would expect to observe $p$ and $q$ values such that $\gamma<0$, indicating that millimeter grains are indeed in the strength regime.

No measurements of $p$ have been made in other debris disks beyond the Kuiper Belt. In this sense, it is currently impossible to compare our measurement of $p$ to the velocity dispersion of bodies in other disks. However, measurements of $q$ have been made for several debris disks. \cite{macgregor2016} used the VLA to observe debris disks around seven nearby stars: HD 377, 49 Ceti, HD 15115, HD 61005, HD 104860, HD 141569, and AU Mic, and determined their grain size distribution indices $q$. After supplementing this sample with other disk observations, they derived a weighted mean for $q = 3.36 \pm 0.02$. Some of these systems have inclinations that are nearly edge-on: HD 377 has an inclination $i = 85 \pm 5\degree$ \citep{choquet2016}, HD 15115 has an inclination $i = 86.3 \pm 0.3\degree$ \citep[][]{sai2015}, and HD 61005 has an inclination $i = 84.1 \pm 0.2$ \citep[][]{olofsson2016}. These inclinations may be large enough to directly resolve the vertical structure of these disks, and multi-wavelength observations of these disks will enable astronomers to directly compare their collisional cascades with steady-state models and with AU Mic and the Kuiper Belt. 

Other debris disk candidates which are bright and close to edge-on include the disk around $\beta$ Pictoris \citep[e.g.][]{smith1984, matra2019, janson2021} and the recently discovered disk around BD$+$45$\degree$598 \citep{hinkley2021}. If these candidates could be observed at both high resolution, necessary to resolve their vertical structures, and at multiple wavelengths such that the relationship between observed wavelengths and disk aspect ratio is significant, then their collisional cascades could be parameterized in a similar fashion to the process described in this work. The disk around $\beta$ Pictoris is especially fascinating in that its vertical structure is more complicated than the vertical structure of the AU Mic debris disk; for example, \cite{matra2019} modelled the vertical structure of the $\beta$ Pictoris disk's mm-sized grains with two Gaussians, since a single Gaussian was unable to reproduce the vertical distribution of dust within the disk. It is ultimately challenging to totally contextualize these results with respect to other (extrasolar) circumstellar disks, since our measurements are the first of their kind. However, our results provide support for the idea that steady-state collisional cascades may be ongoing in a debris disk beyond our solar system \citep[][]{dohnanyi1969, pan2012}. 

\section{Conclusions}\label{chap6}

This work presents 450\,$\mu$m observations of the AU Microscopii debris disk from ALMA which we use to investigate the collisional physics within the disk. AU Microscopii's debris disk is well suited for a parameterization of its collisional physics \citep[within the framework presented in][]{pan2012} since the disk is viewed edge-on. This quality of the disk enables measurements of its scale height, and hence the velocity dispersion of the grains within the disk. The main findings of this work are presented below:

\begin{enumerate}
    \item The morphology of the 450 $\mu$m emission is well-described by an axisymmetric disk. We do not detect any features consistent with the fast-moving features described by \cite{boccaletti2015}. The vertical structure at 450 $\mu$m is spatially resolved. We simultaneously fit a parametric model of the radial and vertical density structure to both 450\,$\mu$m and 1.3\,mm ALMA data \citep[described in][]{daley2019} which assumes an identical viewing geometry at both wavelengths. The inclination of the disk is measured to be $i = 88.51^{+0.44}_{-0.05}$ degrees. 
    
    \item Between $\lambda =  450$\,$\mu$m and $\lambda =  1.3$\,mm, we find a scale height ratio $h_{1300}/h_{450} = 1.35 \pm 0.09$. Our measurement of the aspect ratio at 1.3 mm is slightly lower but agrees within the uncertainties of the 1.3 mm-only fit by \cite{daley2019}. Nevertheless, our results confirm what was already apparent in the image domain: that the vertical FWHM of the disk is thinner at 450\,$\mu$m than at 1.3\,mm.  
    
    \item The dust opacity index $\beta$ and aspect ratios of the disk can be converted into parameters that describe the collisional cascade of the disk. Per the framework described by \citet{pan2012}, which builds on the work of \citet{dohnanyi1969} by exploring the effects of grain-sizes on steady-state equilibrium, we calculate a velocity dispersion power law index $p = 0.28 \pm 0.06$ by comparing the aspect ratio of the disk height at both wavelengths, and a grain size distribution power law index $q = 3.03 \pm 0.02$ from modeling the dust opacity index $\beta$. These values place AU Mic's collisional cascade within the ``gravity regime'' where collisional damping is caused by catastrophic collisions, and smaller bodies are more easily destroyed than larger ones. This result is surprising, since millimeter-size grains would be expected to be in the strength regime instead.  Possible explanations for the discrepancy include ripples in the grain size distribution (which could plausibly occur at the necessary ``wavelength" in grain size, although the amplitude is expected to be small because of the relatively low influence of radiation pressure in the AU Mic system), or a transport-dominated disk due to the influence of the stellar wind (although the collision rate should be high compared to the radial transport rate). 
    
    \item The measured value of $p$, $q$ for the AU Mic debris disk implies that the velocities of grains traced by the 450 $\mu$m continuum emission are damped by destructive interactions. However, it is difficult to draw more general conclusions; more observations of the disk across a wider range of wavelengths (such as visible light, mid-IR, or longer-$\lambda$ radio) are necessary to confirm whether the disk's collisional cascade is globally governed by catastrophic collisions in the gravity regime. Ultimately, we hope that these surprising results will spark follow-up work both theoretical and observational to elucidate how representative (or not) the AU Mic system is and whether new physical frameworks are necessary to interpret the results.
    
    \item Bright, edge-on debris disks are best suited for investigations into their collisional physics, and future work in parameterizing the collisional cascades of other extrasolar debris disks will eventually allow us to contextualize AU Mic's debris disk and the Kuiper Belt within a broader framework of collisional physics in debris disks. 
    
\end{enumerate}

\section{Acknowledgements}

We thank the anonymous referee for insightful comments that improved the paper.
We thank Sasha Krivov, Grant Kennedy, and Luca Matr\'a for thoughtful discussion. We thank Cail Daley for sharing code and for constructive feedback. DV is funded by an Open Study/Research Award from the Fulbright U.S. Student Program.  AMH and DV gratefully acknowledge support from a Cottrell Scholar Award from the Research Corporation for Science Advancement. MAM acknowledges the National Aeronautics and Space Administration under award number 19-ICAR192-0041.

This paper makes use of the following ALMA data: ADS/JAO.ALMA\#2016.1.00878.S, ADS/JAO.ALMA\#2012.1.00198.S. ALMA is a partnership of ESO (representing its member states), NSF (USA) and NINS (Japan), together with NRC (Canada) and NSC and ASIAA (Taiwan), in cooperation with the Republic of Chile. The Joint ALMA Observatory is operated by ESO, AUI/NRAO and NAOJ. This research has made use of NASA's Astrophysics Data System and Astrophysics Data System Bibliographic Services.

This work has made use of data from the European Space Agency (ESA) mission {\it Gaia} (\url{https://www.cosmos.esa.int/gaia}), processed by the {\it Gaia} Data Processing and Analysis Consortium (DPAC, \url{https://www.cosmos.esa.int/web/gaia/dpac/consortium}). Funding for the DPAC has b`een provided by national institutions, in particular the institutions participating in the {\it Gaia} Multilateral Agreement. This research has made use of the SIMBAD database, operated at CDS, Strasbourg, France. 

This research made use of Astropy, a community-developed core Python package for Astronomy \citep{2018AJ....156..123A, 2013A&A...558A..33A}. This research made use of pandas \citep{McKinney_2010, McKinney_2011}. This research made use of NumPy \citep{harris2020array}. This research made use of \texttt{emcee} \citep[][]{foremanmackey2013}. This research made use of the \texttt{CASA} software package \citep[][]{mcmullin2007}.

\acknowledgements{}

\bibliography{bibs}

\end{document}